\documentclass[
  aps,
  pra,
  amsmath,amssymb,amsfonts,
  twocolumn,
  superscriptaddress,
  longbibliography
]{revtex4-2}
\usepackage{soul}
\usepackage{dsfont}
\usepackage{graphicx}
\usepackage{mathtools}
\usepackage[normalem]{ulem}
\usepackage{dcolumn}
\usepackage{bm}
\usepackage{latexsym}
\usepackage{array}
\usepackage{xcolor}
\usepackage{xurl}
\usepackage{placeins}
\usepackage[caption=false]{subfig}
\usepackage{geometry}
\usepackage[colorlinks=true,linkcolor=blue,urlcolor=blue,citecolor=blue,breaklinks=true,pdfusetitle]{hyperref}
\usepackage[none]{hyphenat}
\newcommand {\id}{\mathds{1}}
\newcommand{\ket}[1]{\left|#1\right\rangle}
\newcommand{\bra}[1]{\langle#1|}

\newcommand{\ketbra}[2]{|#1\rangle\langle#2|}

\begin{document}
\title{\texorpdfstring{$\sigma$-VQE: Excited-state preparation of quantum many-body scars with shallow circuits}{sigma-VQE: Excited-state preparation of quantum many-body scars with shallow circuits}}
\author{Eoin Carolan}
\affiliation{Center for Nonlinear and Complex Systems, Dipartimento di Scienza e Alta Tecnologia,
Universit\`a degli Studi dell’Insubria, via Valleggio 11, 22100 Como, Italy}
\affiliation{Istituto Nazionale di Fisica Nucleare, Sezione di Milano, via Celoria 16, 20133 Milano, Italy}
\author{Nathan Keenan}
\affiliation{Instituto de Física Interdisciplinar y Sistemas Complejos (IFISC), UIB–CSIC
UIB Campus, Palma de Mallorca, E-07122, Spain}

\author{Gabriele Cenedese}
\affiliation{Instituto de Física Interdisciplinar y Sistemas Complejos (IFISC), UIB–CSIC
UIB Campus, Palma de Mallorca, E-07122, Spain}

\author{Giuliano Benenti}
\affiliation{Center for Nonlinear and Complex Systems, Dipartimento di Scienza e Alta Tecnologia,
Universit\`a degli Studi dell’Insubria, via Valleggio 11, 22100 Como, Italy}
\affiliation{Istituto Nazionale di Fisica Nucleare, Sezione di Milano, via Celoria 16, 20133 Milano, Italy}
\begin{abstract}
We present and benchmark a type of variational quantum eigensolver (VQE), which we denote $\sigma$-VQE. It is designed to target mid-spectrum eigenstates and prepare quantum many-body scar states. The approach leverages the fact that noisy intermediate-scale quantum devices are limited in their ability to generate generic highly entangled states. This modified VQE pairs a low-depth circuit with an energy-selective objective that explicitly penalizes energy variance around a chosen target energy. The cost function exploits the limited expressibility of the shallow circuit as atypical low-entanglement eigenstates such as scar states are preferentially selected. We validate this mechanism across two complementary families of models that contain many-body scar states: the Shiraishi–Mori embedding approach and a matrix product state parent Hamiltonian construction. We define an unbiased estimation scheme for the nonlinear cost function that is compatible with qubit-wise commuting grouping and bitstring reuse. A proof-of-principle demonstration using a small-system instance was performed on IBM Fez (Heron r2 QPU). These results motivate its use as a practical algorithm for detecting quantum many-body scars and variationally generating states with appreciable scar state overlap.
\end{abstract}
\date{\today}

\maketitle
\section{Introduction}
Variational methods for approximating eigenstates date back more than a century \cite{ritz1909neue}. The same principles underpin a modern hybrid classical-quantum counterpart — the variational quantum eigensolver (VQE) \cite{tilly2022variational}. The VQE has become a standard paradigm for noisy intermediate-scale quantum (NISQ) hardware \cite{preskill2018quantum,eisert2025mind}, with applications ranging from quantum chemistry \cite{fonseca2025introduction,hu2022benchmarking}, to materials science \cite{yoshioka2022variational,stanisic2022observing}, and quantum machine learning \cite{huijgen2024training}. In an iteration of the VQE algorithm, a parametrized circuit prepares a trial state, a cost (or objective) function is estimated by measurements and classical post-processing, and a classical optimizer updates the circuit parameters to ideally lower the cost function for the next iteration. It is one example of a hybrid quantum-classical algorithm \cite{mcclean2016theory,cerezo2021variational}. While textbook VQEs focus on ground states by minimizing $\langle H\rangle$ with respect to the circuit state, many research questions that probe excited-state structure, finite-temperature physics, and nonequilibrium initial-state engineering demand methods that can target interior eigenstates without requiring deep circuits or full diagonalization.

Quantum many-body scars (QMBS) are atypical eigenstates that violate the eigenstate thermalization hypothesis (ETH) in a spectrum that otherwise follows it  \cite{deutsch1991quantum,srednicki1994chaos,d2016quantum}. The ETH predicts that individual eigenstates at fixed energy density in generic interacting systems reproduce microcanonical thermodynamics for local observables. Scarred eigenstates are counterexamples \cite{bernien2017probing,turner2018weak}. They exhibit anomalously low entanglement compared to nearby states, which enables long-lived coherent dynamics and revivals from simple product-state initial conditions. Although towers of scar states and strong revivals are especially prominent in parts of the scar literature, they are not necessary for the terminology used here, which follows the standard usage that also encompasses exact embedded Shiraishi–Mori-type constructions \cite{shiraishi2017systematic}. These features make QMBSs relevant testbeds for exploring how nonergodic structure can persist without integrability or disorder. They have potential practical applications due to their ability to stabilize coherent dynamics at energies where typical states are highly entangled and rapidly thermalize. For recent reviews on QMBSs, see Refs. \cite{chandran2023quantum,moudgalya2022quantum}.

In this work we modify the standard VQE to target QMBSs by using an eigenstate-targeting objective that is naturally suited to mid-spectrum search \cite{Cenedese2025}. We denote this protocol the $\sigma$-VQE. The central modification relative to the ``textbook" implementation of the VQE is a nonlinear cost function that depends on the first and second moments of the Hamiltonian and explicitly penalizes energy variance at a chosen target energy $E_{\mathrm{tar}}$. By construction, the objective is minimized by eigenstates with energy closest to $E_{\mathrm{tar}}$, while coherent superpositions of eigenstates incur a variance penalty. This principle is closely related to a wider family of excited-state VQE strategies explored in other contexts \cite{CadiTaziThom2024FoldedSpectrumVQE,Zhang2021AdaptiveVQEX,higgott2019variational,nakanishi2019subspace,mcclean2017hybrid,gocho2023excited}, but here it is deployed alongside shallow quantum circuits specifically as a diagnostic and preparation primitive for QMBSs. The same cost function was introduced in \cite{Cenedese2025} under the name VQE-S and studied in ideal statevector simulations. Here we develop its finite-shot unbiased implementation, extend the benchmarking to two complementary scar constructions including a tunable entanglement parent Hamiltonian setting, and provide a proof-of-principle hardware demonstration on IBM Fez.

A second key design choice is to combine the $\sigma$-VQE cost function with a deliberately shallow circuit ansatz \cite{sim2019expressibility}. Shallow circuits are often treated as a limitation of NISQ devices, while here they become an asset. Generic mid-spectrum eigenstates typically exhibit volume-law entanglement, whereas scarred eigenstates atypically display low entanglement. A low-depth circuit strongly constrains the entanglement accessible to prepared states. In the benchmark settings studied here, that representational restriction disfavors generic highly entangled mid-spectrum eigenstates while leaving the anomalous low-entanglement target states accessible. We stress, however, that this is a statement about representational bias rather than a general claim that shallow circuits are always easier to optimize. The low-depth ansatz has the additional benefit of exposing the system to less noise than would be seen for deeper circuits.

A practical challenge shared by VQE-style algorithms is circuit overhead. Estimating Hamiltonian-based objective functions can involve measuring many Pauli strings and thus preparing the same circuit many times over. This motivates the use of estimators and measurement strategies that (a) exploit operator structure and grouping, (b) remain unbiased at finite shot count, and (c) make maximal use of each bitstring collected by the measurement of the system at the end of the circuit. In the $\sigma$-VQE implementation presented in this work we address these requirements by combining a qubit-wise commuting (QWC) measurement strategy with an estimator designed to reuse each sampled bitstring across many strings within a commuting group, together with an unbiased construction for the nonlinear terms. Further improvement of the algorithm can naturally be found in the broader landscape of measurement-frugal estimation schemes (including grouping and derandomization) that aim to reduce the number of circuit executions required for a fixed estimation accuracy \cite{verteletskyi2020measurement,huang2020predicting,huang2021efficient}. 
We validate the experimental feasibility of $\sigma$--VQE through a proof-of-principle realization on superconducting hardware using an IBM Heron r2 processor.

The structure of the paper is as follows. In Sec. \ref{sec:vqe}, we introduce the $\sigma$-VQE algorithm. We specify the shallow entangling ansatz, the cost function, and the two classical optimizers we consider. We use a gradient-based optimizer (ADAM) in the noiseless setting, where gradients (needed to update circuit parameters iteration-to-iteration) can be estimated using parameter-shift rules \cite{Mitarai2018QuantumCircuitLearning,Schuld2019AnalyticGradients}, and a gradient-free optimizer (SPSA) when contending with device noise and significant shot budget limitations. We then present our QWC-compatible measurement scheme and construct an unbiased finite-shot estimator for the nonlinear objective. In Sec. \ref{sec:scars} we outline two families of benchmark Hamiltonians that contain scarred eigenstates. The first is the Shiraishi–Mori projector-embedding construction \cite{shiraishi2017systematic} to embed a product eigenstate at finite energy density into an otherwise chaotic spectrum. The second uses a parent Hamiltonian embedding of random matrix product states (MPSs) \cite{larsen2024phase}, enabling controlled tuning of scar state entanglement. Finally, in Sec. \ref{sec:results} we benchmark $\sigma$-VQE performance in idealized simulation and under realistic finite-shot and hardware noise, including a proof-of-principle execution on superconducting quantum hardware.

\section{\texorpdfstring{$\sigma$}{sigma}-VQE}\label{sec:vqe}
The $\sigma$-VQE is a hybrid quantum-classical variational algorithm. In the ideal case a noiseless parametrized quantum circuit $U(\theta)$ prepares the state $\ket{\psi(\theta)}$. Expectation values for observables relevant for a cost objective (typically the Hamiltonian) are estimated on that state and a classical optimizer uses this information to update the circuit parameters $\theta$ to find a parameter configuration that lowers the cost function. This process repeats until convergence to an optimal solution (within an acceptable tolerance) or resource limits are reached. The key differences with $\sigma$-VQE relative to textbook ground-state VQE concern the cost function and the way in which we estimate it. The cost is chosen to select an eigenstate of our Hamiltonian of interest with energy close to a specified target energy, $E_\mathrm{tar}$. The estimator is built to remain unbiased under finite-shot sampling, and to be compatible with qubit-wise commuting (QWC) grouping. The optimizer can be either a gradient-based routine such as ADAM or a gradient-free stochastic routine such as SPSA. We summarize all these elements below.

\subsection{Circuit ansatz and cost function}
\begin{figure}[hbt!]
    \centering
    \includegraphics[width=0.45\textwidth]{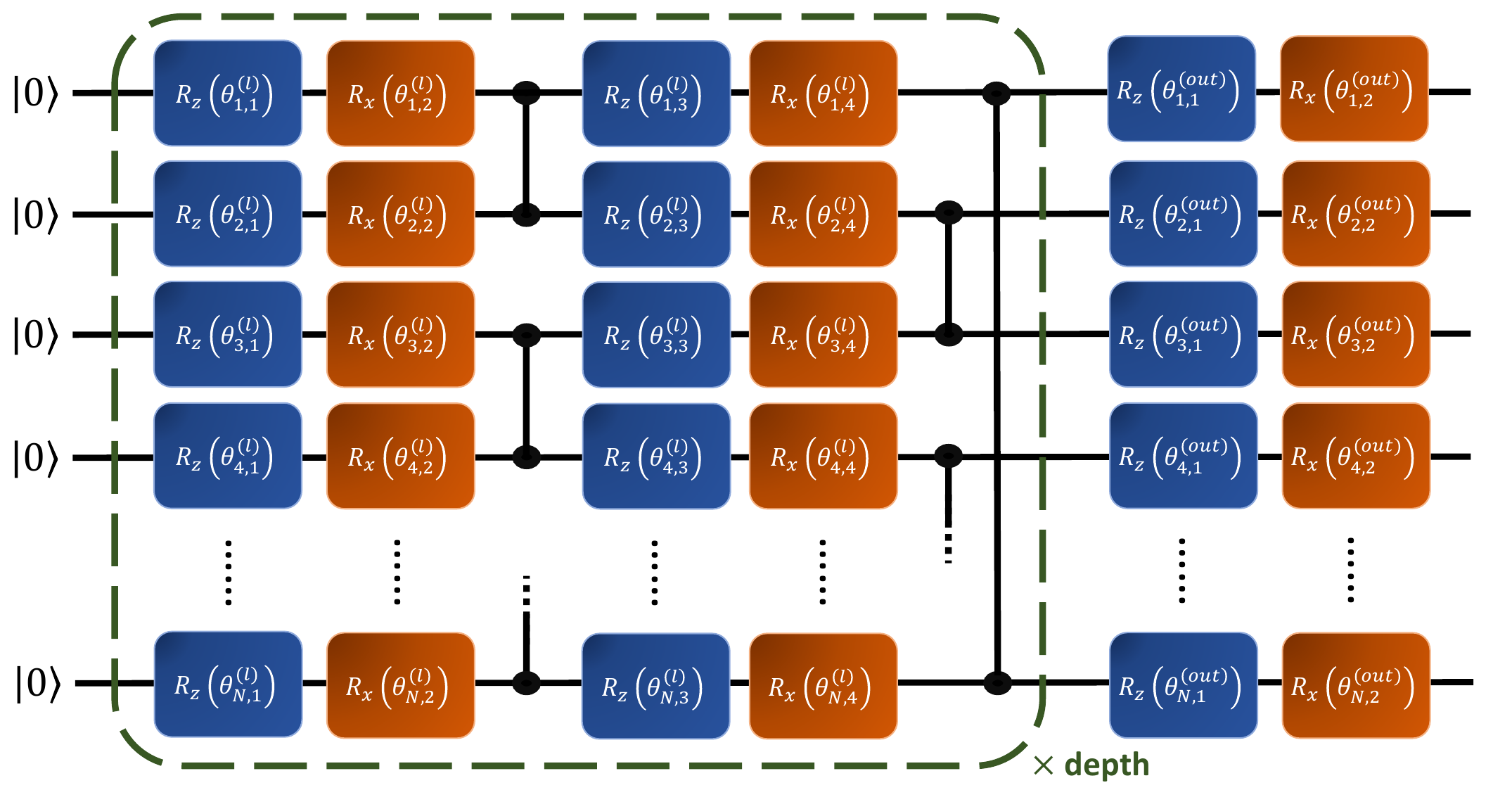} 
    \caption{Hardware-efficient ansatz used throughout this work. $N$ is the number of qubits in the circuit, and the index $(l)$ runs from 1 to the total depth of the ansatz. Each layer applies single-qubit rotations on all qubits followed by a pattern of two-qubit entangling gates (CZ) on a periodic nearest-neighbor chain. The circuit depth denotes the number of such layers.}
    \label{fig:HE}
\end{figure}

The circuit ansatz is a key element of VQEs that must balance expressibility with efficiency for the learning task \cite{kim2021universal}. We choose a hardware-efficient ansatz built from $d$ layers of single-qubit rotations followed by nearest-neighbor CZ entangling gates, as shown in Fig. \ref{fig:HE}. The variational parameters $\theta$ are the collection of rotation angles in the one-qubit gates, while the CZ gates are fixed and parameter-free. The circuit is initialized in the all-zero $N$-qubit computational basis state, $\ket{0}^{\otimes N}$. The hardware-efficient ansatz is typical in NISQ-based VQE chemistry and materials applications \cite{kandala2017hardware}. It is designed to be relatively agnostic to the specific Hamiltonian one wishes to implement, and instead be relatively easy to implement on a device using native operations. We intentionally restrict the circuit depth so that we preferentially target low-entanglement states. This expressive bias is central to using $\sigma$-VQE as a scar-hunting primitive, since generic mid-spectrum eigenstates are typically volume-law entangled while scars are not \cite{Cenedese2025, PhysRevB.107.024204}. 

The cost function that defines the $\sigma$-VQE \cite{Cenedese2025} is designed to penalize both deviation from a target energy and non-eigenstates. Let
\begin{equation}
    \langle H\rangle_{\theta}=\bra{\psi(\theta)}H\ket{\psi(\theta)},\quad \langle H^2\rangle_{\theta}=\bra{\psi(\theta)}H^2\ket{\psi(\theta)}.
\end{equation}
We then define the cost function as
\begin{equation}\label{eq:cost}
C(\theta) = a \langle(H - E_{\mathrm{tar}})^2\rangle_{\theta} + b \left(\langle H^2\rangle_{\theta} - \langle H\rangle_{\theta}^2\right),
\end{equation}
where $\theta$ denotes the vector of variational parameters, $E_{\mathrm{tar}}$ is the target energy from which deviations in the energy of the circuit state are penalized, and $a$ and $b$ are real positive weights where we take $a+b=1$ without loss of generality. The term weighted by $a$ forces the mean energy toward the target value $E_{\mathrm{tar}}$. The term weighted by $b$ is the energy variance. If $\ket{\psi(\theta)}$ is an exact eigenstate of $H$ with eigenvalue $\lambda$, then $\langle H\rangle_{\theta}^2=\langle H^2\rangle_{\theta}=\lambda^2$, and 
\begin{equation}
    C(\theta)=a(\lambda-E_{\mathrm{tar}})^2.
\end{equation}
This highlights the importance of the squaring of the first term to minimize $C(\theta)$ for eigenstates at $\lambda=E_{\mathrm{tar}}$: any state that is a coherent superposition of different eigenstates incurs a positive variance penalty. We fix $a+b=1$ as only their relative weighting is meaningful, while an overall common rescaling would simply rescale the cost function. We therefore use $a=b=1/2$ as a simple symmetric benchmark choice rather than as a finely tuned optimum. Unless otherwise stated we set $E_{\mathrm{tar}}=0$, which matches the embedded scar energy in our testbed models which we define later in Sec. \ref{sec:scars}.
\\
\subsection{Gradient-based optimizer: ADAM}
There is a choice over what classical optimization scheme is used in hybrid variational quantum algorithms \cite{ComparativeOptimizersVQE2025, Lavrijsen2020ClassicalOptimizers}. Unless restricted by noise or resources, we use ADAM (short for Adaptive
Moment Estimation) as it has been shown to be particularly robust, and typically converges quickly. ADAM is a stochastic gradient method which utilizes adaptive first and second moment estimates of the gradient. For our purposes we estimate the gradient for each parameter using the parameter-shift rule. Let $\theta_t$ be the parameter vector at iteration $t$. Let $g_t=\nabla_{\theta}C(\theta_t)$ be the gradient of the cost function, Eq. \ref{eq:cost}. ADAM maintains two auxiliary vectors: a running exponential average of gradients, and a running exponential average of squared gradients (or velocity)
\begin{align}
m_t &= \beta_1 m_{t-1} + (1 - \beta_1) g_t,\\
v_t &= \beta_2 v_{t-1} + (1 - \beta_2) g_t\odot g_t ,
\end{align}
where $\odot$ denotes the Hadamard product. We take the hyperparameters $\beta_1$ and $\beta_2$ to be $0.9$ and $0.999$, respectively. Then we correct for bias as both $m_0$ and $v_0$ are $0$. Bias-corrected estimates are then
\begin{equation}
    \hat{m}_t = \frac{m_t}{1 - \beta_1^t}, \quad \hat{v}_t = \frac{v_t}{1 - \beta_2^t}.
\end{equation}
The update rule for each component is
\begin{equation}
    \theta_{t+1,k} = \theta_{t,k} - \alpha \frac{\hat{m}_{t,k}}{\sqrt{\hat{v}_{t,k}} + \epsilon},
\end{equation}
where $\alpha$ is the learning rate, and $\epsilon$ is a numerical stabilization hyperparameter which we take to be $10^{-8}$. To update the parameters we need to estimate the gradient $g_t$. As the circuit consists of fixed two-qubit entangling gates and single-qubit rotations of the form $\exp(-i\theta_kG_k/2)$, where $G_k$ is a Pauli operator (or tensor product thereof), we make use of the parameter-shift rule (PSR) to calculate the gradient. For an observable $O$, the PSR implies that the derivative $\partial_{\theta_k}\langle O\rangle_{\theta}$ can be computed as
\begin{equation}\label{eq:PSR}
 \partial_{\theta_k}\langle O\rangle_{\theta}= \frac{1}{2} \left[\langle O\rangle_{\theta+\frac{\pi}{2}}-\langle O\rangle_{\theta-\frac{\pi}{2}}\right],
\end{equation}
where only parameter $\theta_k$ is shifted by $\pm \pi/2$. As the cost function for $\sigma$-VQE is a linear combination of $\langle H\rangle_{\theta}$, $\langle H^2\rangle_{\theta}$, and $\langle H\rangle_{\theta}^2$, we cannot simply apply the PSR directly to the cost function as in the case of a ground state search. Instead we have
\begin{equation}
    \partial_{\theta_k}C=(a+b) \partial_{\theta_k}\langle H^2\rangle_{\theta}-(2aE_{\mathrm{tar}}+2b\langle H\rangle_{\theta})\partial_{\theta_k}\langle H\rangle_{\theta}.
\end{equation}
 Both $ \partial_{\theta_k}\langle H^2\rangle_{\theta}$ and $\partial_{\theta_k}\langle H\rangle_{\theta}$ can be evaluated via \eqref{eq:PSR}. The full gradient can then be assembled in an unbiased way by using the PSR for each derivative, with an additional evaluation of $\langle H\rangle_{\theta}$. 
 
This method of calculating the gradient is exact, but the calculation of the gradient for each optimization iteration requires $2P+1$ estimates of expectation values, where $P$ denotes the number of parameters and we assume both $ \langle H\rangle$ and $ \langle H^2\rangle$ can be estimated simultaneously from the same circuit measurement (more on which later). 

\subsection{Gradient-free optimizer: SPSA}
Expectation values -- and therefore also the cost function -- \eqref{eq:cost} are estimated from a finite number of shots (circuit preparations and executions) on quantum hardware. Using the PSR to calculate the gradient demands $2P+1$ expectation values to be estimated. We can avoid explicit PSR gradients by optimizing with a stochastic gradient approximation. This will prove useful in cases where our estimation of the cost function is particularly noisy, or when the number of shots is severely limited. The gradient-free method we will use is Simultaneous Perturbation Stochastic Approximation (SPSA) \cite{spall2002implementation}. SPSA replaces the gradient by a simultaneous two-point finite difference along a random direction in parameter space.\\

We now outline the SPSA method. At iteration $t$ of the optimization procedure we draw a random perturbation vector $\Delta_t$. Each component $\Delta_{t,k}$ is independently and randomly drawn from $\{-1,1\}$. SPSA then evaluates the cost function at two shifted parameter vectors
\begin{equation}
    \theta_t^{(+)}=\theta_t+c_t\Delta_t,\quad \theta_t^{(-)}=\theta_t-c_t\Delta_t,
\end{equation}
where $c_t=c_0(t+1)^{-\gamma}$ is a decay hyperparameter. The finite-difference approximation of the gradient is given by 
\begin{equation}\label{eq:shifted}
    \hat{g}_{t,k}=\frac{C(\theta_t^{(+)})-C(\theta_t^{(-)})}{2c_t\Delta_{t,k}}.
\end{equation}
The advantage of SPSA is that this estimate requires only $2$ cost evaluations per iteration (instead of $2P+1$), and therefore is independent of the number of parameters. We then update the parameters as
\begin{equation}
    \theta_{t+1}=\theta_t-a_t\hat{g}_{t},
\end{equation}
where
\begin{equation}
    a_t=\frac{a_0}{(A+t+1)^{\alpha}}.
\end{equation}
The algorithm wraps each updated angle back to the interval $(-\pi,\pi]$. SPSA is well suited to NISQ devices as it is known to tolerate device noise and demands a lower shot overhead \cite{kandala2017hardware,pellow2021comparison}. We have a number of hyperparameters, $\{a_0,c_0,A,\gamma,\alpha\}$, and unlike ADAM, SPSA is quite sensitive to their choice. When utilizing SPSA we optimize the hyperparameters using the methods prescribed in \cite{spall2002implementation}.

\subsection{Unbiased estimation of the cost function}\label{subsec:unbiased}

We now describe how to estimate the expectation values $\langle H\rangle_{\theta}$, $\langle H^2\rangle_{\theta}$, and $\langle H\rangle_{\theta}^2$ needed for both PSR and SPSA methods from finite shots in a way that is unbiased (with respect to shot noise), and allows for simultaneous use of circuit measurement outcomes for each value.

We work with $N$ qubits. We denote $\mathcal{P}=\{\id,\sigma_x,\sigma_y,\sigma_z\}^{\otimes N}$ as the set of Pauli strings. We expand the operators needed for the cost function in the Pauli string basis as
\begin{equation}
H \;=\;\alpha_{\id}\id+\sum_{i} \alpha_i P_i,
\qquad
H^2 \;=\; \beta_{\id}\id+\sum_{i} \beta_i P_i,
\label{eq:expansions}
\end{equation}
with real coefficients $\alpha_i,\beta_i$. For a given circuit state $\ket{\psi(\theta)}$, the expectation values we need are
\begin{equation}
    \langle H\rangle_{\theta}=\alpha_{\id}+\sum_{i} \alpha_i \langle P_i\rangle_{\theta},\quad\langle H^2\rangle_{\theta}=\beta_{\id}+\sum_{i} \beta_i \langle P_i\rangle_{\theta}.
\end{equation}
Measuring $\langle P_i\rangle_{\theta}$ for all strings separately is inefficient and wastes information that can be shared across shots. Instead we exploit qubit-wise commuting (QWC) structure \cite{yen2020measuring,dalfavero2023k} and use a classical shadow-motivated approach to simultaneous estimation of expectation values \cite{huang2020predicting}. A Pauli string $P_i$ is compatible with a measurement basis $B\in\{\sigma_x,\sigma_y,\sigma_z\}^{\otimes N}$ if on each qubit $P_i$ either acts as the identity, or $P_i$ and $B$ act with the same Pauli axis. We define the coverage indicator for this as
\begin{equation}
\delta_B(i)=
\begin{cases}
1, & P_i \text{ covered by } B,\\
0, & \text{otherwise.}
\end{cases}
\end{equation}
Let $\mathcal{B}$ be the set of all bases that cover at least one Pauli string $P_i$ which has nonzero $\alpha_i$ or $\beta_i$. To measure in a basis $B$, local pre-rotations $U_B=\bigotimes_j U_{B_j}$ are applied to each qubit in the register after the circuit is executed. This maps measurement in the computational basis to a measurement in basis $B$. This mapping is done by $U_{\sigma_x}=H$, $U_{\sigma_y}=R_x(\pi/2)$, $U_{\sigma_z}=I$.
The bitstring $q\in\{0,1\}^N$ is then drawn by measurements on each qubit in the computational basis. The distribution of outcome bitstrings is then
\begin{equation}\label{eq:pq}
p_q\equiv\Pr(q\,|\,B,\theta)=\big|\langle q|U_B|\psi_{\theta}\rangle\big|^2.
\end{equation}
For any $i$ covered by $B$, the eigenvalue of $P_i$ on the outcome $q$ is
\begin{equation}
m_i(q)=\prod_{j\in \mathrm{supp}(P_i)} (-1)^{q_j}.
\end{equation}
where $\mathrm{supp}(P_i)$ means we consider only those elements of the bitstring that correspond to non-identity elements of the Pauli string $P_i$. By construction $\mathbb{E}[\,m_i(q)\,|\,B,\theta\,]=\langle P_i\rangle_{\theta}$. Given a fixed budget of circuit executions we allocate shots across measurement bases using importance sampling. We define basis weights
\begin{equation}
w_B^{(H)}=\sum_{i}\! |\alpha_i|\,\delta_B(i),\qquad
w_B^{(H^2)}=\sum_{i}\! |\beta_i|\,\delta_B(i),
\end{equation}
and sample a measurement basis $B$ with
\begin{equation}\label{eq:basisprob}
p_B \;=\; \frac{w_B^{(H)}+w_B^{(H^2)}}{\sum_{B'\in\mathcal{B}}\big(w_{B'}^{(H)}+w_{B'}^{(H^2)}\big)}.
\end{equation}
The coverage probability for a string $i$ is given by 
\begin{equation}
\xi(i)=\sum_{B\in\mathcal{B}} p_B\,\delta_B(i).
\end{equation}
This is the probability that a random basis draw can measure $P_i$. After $S$ shots, where shot $s=1,2,...,S$ has corresponding measurement setting and outcome $(B^{(s)},q^{(s)})$, we form the unbiased estimators 
\begin{align}\label{eq:unbias}
    \widehat{\langle H\rangle}
&= \alpha_{\id} + \frac{1}{S}\sum_{s=1}^S \sum_{i}\frac{\alpha_i\,\delta_{B^{(s)}}(i)}{\xi(i)}\,m_i\big(q^{(s)}\big),
\qquad\\
\widehat{\langle H^2\rangle}
&= \beta_{\id} + \frac{1}{S}\sum_{s=1}^S \sum_{i}\frac{\beta_i\,\delta_{B^{(s)}}(i)}{\xi(i)}\,m_i\big(q^{(s)}\big).
\end{align}
These estimators are unbiased because each Pauli-string contribution is reweighted by the inverse of its sampling probability, $\xi(i)$. In addition to reusing each shot across all covered strings, the importance sampling additionally concentrates shots on bases that cover strings with large $|\alpha_i|$ and $|\beta_i|$ which reduces variance for fixed $S$.

The cost function, Eq. \eqref{eq:cost}, also relies on estimating $\langle H\rangle_{\theta}^2$. If we naively square our estimator $ \widehat{\langle H\rangle}$ we will introduce bias at finite S. To avoid this bias we build a U-statistic for $\langle H\rangle_{\theta}^2$. We isolate the per-shot contribution from \eqref{eq:unbias}
\begin{equation}
    Y^{(s)}=\sum_{i}\frac{\alpha_i\,\delta_{B^{(s)}}(i)}{\xi(i)}\,m_i\big(q^{(s)}\big).
\end{equation}
The unbiased estimator is then constructed as
\begin{equation}
     \widehat{\langle H\rangle^2}=\frac{(\sum_sY^{(s)})^2-\sum_s(Y^{(s)})^2}{S(S-1)}.
\end{equation}
It satisfies $\mathbb{E}[ \widehat{\langle H\rangle^2}]=\langle H\rangle_{\theta}^2$ provided that the shots are independent. Our estimator for the cost function is then
\begin{equation}
\widehat{C}=(a+b)\widehat{\langle H^2\rangle}-2aE_{\mathrm{tar}}\widehat{\langle H\rangle}-b\widehat{\langle H\rangle^2}+aE_{\mathrm{tar}}^2.
\label{eq:chatfull}
\end{equation}
Unless otherwise stated, we take $E_{\mathrm{tar}}=0$, and $a=b=1/2$, which gives the cost function the following form
\begin{equation}
\widehat{C}=\widehat{\langle H^2\rangle}-\frac{1}{2}\widehat{\langle H\rangle^2}.
\label{eq:chat}
\end{equation}
This estimator is what is queried when we need to estimate the full cost function such as for SPSA. We note that ADAM with the PSR only requires estimation of $\langle H\rangle_{\theta}$  and $\langle H^2\rangle_{\theta}$, with the U-statistic only needed for logging an estimate of the cost.
\subsection{Simulating shot noise}\label{subsec:noise}
Simulating \(S\) independent circuits per optimization iteration to capture
the effect of shot noise is expensive in classical benchmarks, particularly
when we simulate a density matrix to capture hardware noise. We avoid this
by drawing the shot statistics directly. After choosing an average number of
shots per QWC measurement circuit, \(\bar S_{\rm circ}\), we set the total shot
budget for one estimator evaluation at fixed circuit parameters \(\theta\) to
\[
    S = G\bar S_{\rm circ},
\]
where \(G=|\mathcal B|\) is the number of QWC measurement bases. We then
allocate this total budget across bases by sampling a multinomial distribution
\cite{ross2023first} with probabilities \(p_B\) given by Eq.~\eqref{eq:basisprob},
i.e., the distribution of counts obtained from \(S\) independent draws over the
possible measurement bases,
\begin{equation}
    \vec K \sim \mathrm{Mult}(S,\vec p_B),\quad
    \sum_{B\in\mathcal B}K_B=S.
\end{equation}
This allocation gives, on average, more shots to measurement bases covering Pauli strings with larger coefficients in $H$ and $H^2$. Then, for each basis, we draw the histogram of $K_B$ total outcome bitstrings in that basis from the multinomial distribution with probabilities $\vec p_q$,
\begin{equation}
    \vec M^{(B)}\sim \mathrm{Mult}(K_B,\vec p_q),\quad
    \sum_{q\in\{0,1\}^N} M^{(B)}_q=K_B,
\end{equation}
where $p_q$ is given by Eq.~\eqref{eq:pq}. With this we can reconstruct the expectation values using our unbiased estimators as before. This approach samples the same shot and basis-allocation noise as $S$ independent circuit executions, without explicitly simulating each execution individually.

\section{Models with embedded scar states}\label{sec:scars}
We now introduce the Hamiltonians used to benchmark the $\sigma$-VQE. The guiding principle is to construct interacting spin models that host at least one eigenstate with anomalously low entanglement and otherwise display Wigner-Dyson spectral statistics. We consider two constructions. The first is a Shiraishi-Mori type projector embedding of a product scar. The second is a parent Hamiltonian construction for an MPS scar of tunable bond dimension.
\begin{figure*}[t]
    \centering

    \includegraphics[width=0.31\textwidth]{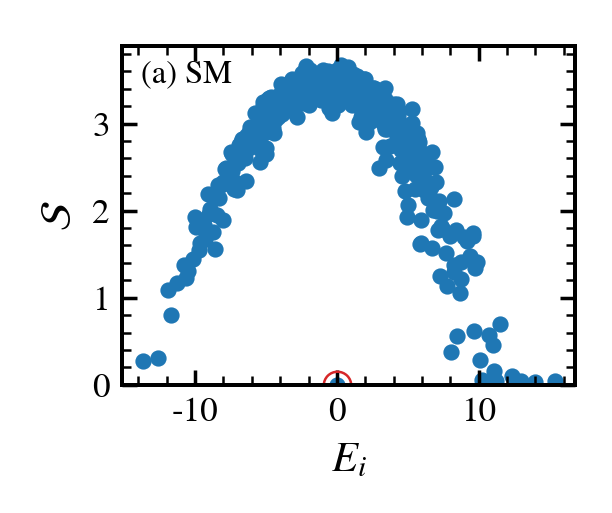}%
    \hspace{0.02\textwidth}%
    \includegraphics[width=0.31\textwidth]{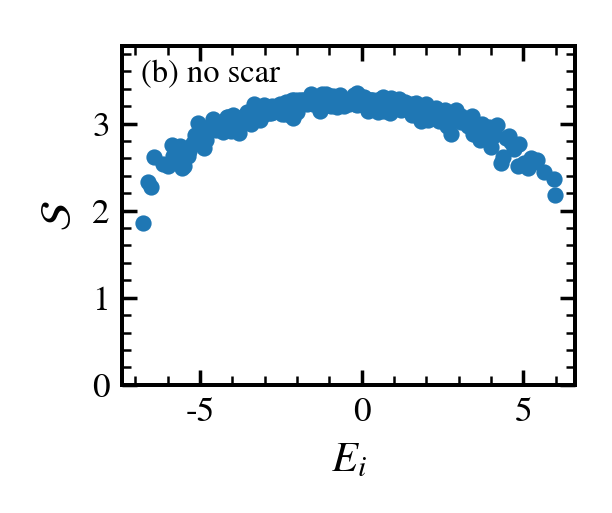}%
    \hspace{0.02\textwidth}%
    \includegraphics[width=0.31\textwidth]{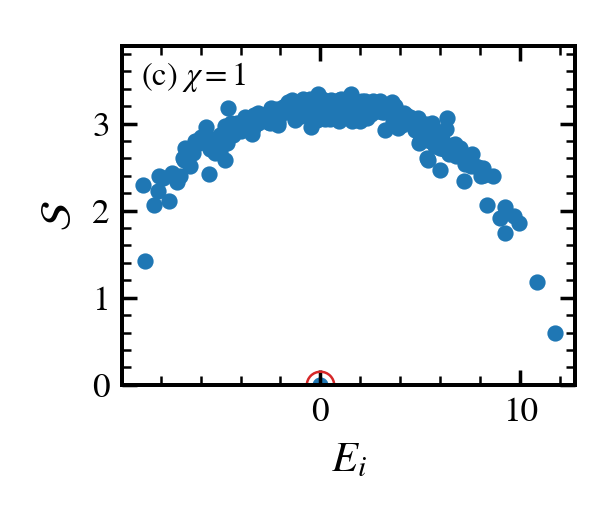}%

    \vspace{0.035\textwidth}

    \includegraphics[width=0.31\textwidth]{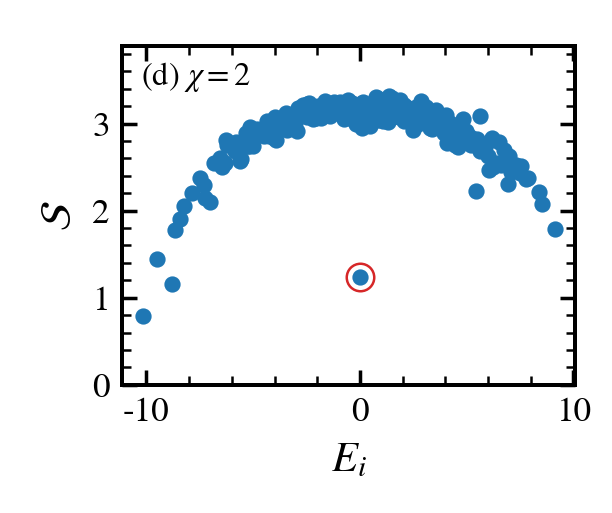}%
    \hspace{0.04\textwidth}%
    \includegraphics[width=0.31\textwidth]{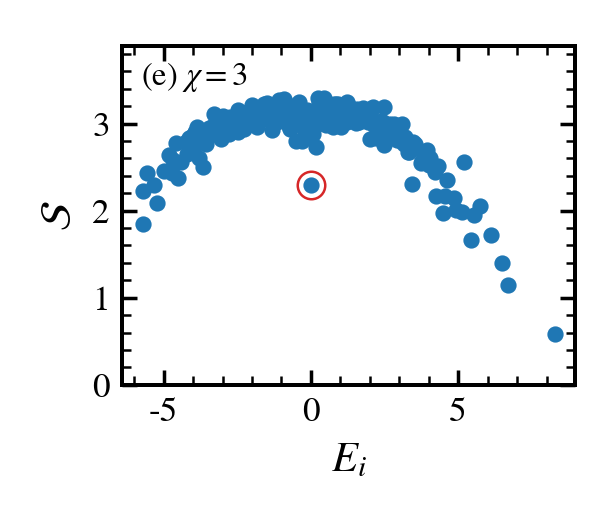}%

    \caption{%
    Bipartite entanglement entropy $\mathcal{S}$ of eigenstates for the benchmark Hamiltonians. The scar states are circled in red. (a) The $N=9$ Shiraishi--Mori (SM) embedding yields a product-state scar at $E=0$. (b) The $N=8$ parent Hamiltonian construction without an embedded scar. (c)--(e) The $N=8$ parent Hamiltonian construction with an embedded MPS scar of bond dimension $\chi=1,2,3$, respectively. Increasing $\chi$ increases the scar entanglement and moves the scar state closer to the thermal band.%
    }
    \label{fig:ent}
\end{figure*}
\subsection{Shiraishi-Mori embedding}
The projector-embedding approach of Shiraishi and Mori \cite{shiraishi2017systematic} is a general method for embedding many-body scar states into the spectrum of a Hamiltonian. It can be readily extended to arbitrary dimensions and model geometries, though we restrict to 2-local interactions on an open spin chain. Let $\langle i,j\rangle$ denote a pair of neighboring sites i and j. Let $h_{ij}$ be a two-spin Hermitian operator and $P_{ij}$ a projector acting on sites $i$ and $j$. The Shiraishi-Mori Hamiltonian is then
\begin{equation}
    H=\sum_{\langle i,j\rangle}P_{ij} h_{ij}P_{ij},
\end{equation}
where $\langle i,j\rangle$ indicates we are summing over all bonds in the model. We independently draw a Haar-random normalized single-qubit state $\ket{\phi_i}$ for each site $i$. The projector for each bond is then
\begin{equation}
    P_{ij}=\id\otimes\id-\ketbra{\phi_i}{\phi_i}\otimes\ketbra{\phi_j}{\phi_j}
\end{equation}
This embeds the product state $\ket{\psi_\mathrm{scar}}=\otimes_{i}\ket{\phi_i}$ as a scar state with eigenvalue 0, since it is annihilated by each $P_{ij}$. The state $\ket{\psi_\mathrm{scar}}$ is clearly a product state, so its bipartite entanglement entropy is strictly zero. We choose the local Hamiltonian terms to be XXZ-type with a symmetry breaking transverse field as in \cite{burke2025taking}
\begin{equation}\label{eq:smxxz}
    h=J(\sigma_x\otimes\sigma_x+\sigma_y\otimes\sigma_y)+\Delta \sigma_z\otimes\sigma_z+b(\sigma_x\otimes\id+\id\otimes\sigma_x),
\end{equation}
with $J=b=1$ and $\Delta=0.7$. We therefore obtain a model that has a product eigenstate with energy zero, while the rest of the spectrum displays Wigner-Dyson level spacing statistics, specifically GOE. This setting is ideal for testing whether $\sigma$-VQE can selectively prepare the scarred eigenstate over the large number of other nearby eigenstates. 

One limitation of the Shiraishi–Mori embedding used here is that the embedded scar is a product state. In numerical and analytical studies of many-body scar phenomenology this is often sufficient, but many physical scar states such as in Rydberg chains \cite{bernien2017probing,turner2018quantum,dooley2023entanglement, hudomal2020quantum} exhibit nontrivial internal entanglement. We therefore introduce a second construction that embeds a target state with tunable entanglement. 
\subsection{Parent Hamiltonian embedding}\label{sec:ph}
We next construct Hamiltonians that host an entangled scar eigenstate whose entanglement structure we can tune. We do this by building a parent Hamiltonian for a chosen matrix product state (MPS). We consider a ring of $N$ qubits. We define the normalized, translationally-invariant MPS state with bond dimension $\chi$ as
\begin{equation}
\ket{\psi_\mathrm{MPS}}=C\sum_{s_1,...,s_N}\mathrm{Tr}[A^{s_1}A^{s_2}...A^{s_N}]\ket{s_1s_2...s_N},
\end{equation}
where $s_i\in[0,1]$, $C$ is a normalization constant, and $A^0,A^1\in\mathbb{C}^{\chi^2}$. The bond dimension $\chi$ controls the amount of entanglement in the MPS state. We construct a Hamiltonian operator that annihilates $\ket{\psi_\mathrm{MPS}}$. We specify an interaction range $D$ for the ring of spins. We then tile the ring with $N$ overlapping length-$D$ blocks. On each block we consider a subspace $\mathcal{A}$ that supports the reduced density matrix of $\ket{\psi_\mathrm{MPS}}$ on those $D$ sites. The orthogonal complement $\mathcal{A}_C$ within the full $2^D$-dimensional Hilbert space of the block is the set of local states that we will build the Hamiltonian from. We require
\begin{equation}
    2^D>\chi^2
\end{equation}
in order for the complement space to be non-empty. Let $\{\psi_n\}$ be an orthonormal basis for $\mathcal{A}_C$. We define a $D$-local Hermitian operator akin to an annihilation operator acting on the $i^{th}$ block as
\begin{equation}
h_i=\sum_{m,n\in\mathcal{A}_C}c_{mn}\ketbra{\psi_m}{\psi_n}.
\end{equation}
As the MPS state has no support in the complement space, a projector operator $h_i$ constructed from states from $\mathcal{A}_C$ will annihilate the MPS state. We can choose $c_{mn}$ to be positive semidefinite. Each $h_i$ is then positive semidefinite along with satisfying $h_{i} \ket{\psi_\mathrm{MPS}}=0$. The parent Hamiltonian
\begin{equation}
    H_P=\sum_{i=1}^Nh_i,
\end{equation}
will then have $\ket{\psi_\mathrm{MPS}}$ as its ground state. However, we are interested in mid-spectrum scars, not ground states. To embed $\ket{\psi_\mathrm{MPS}}$ into the interior of the spectrum we instead choose coefficients $c_{mn}$ that are not all positive. A simple choice is \cite{larsen2024phase}
\begin{equation}
    c_{nm}=(-1)^n\delta_{nm},
\end{equation}
which alternates signs along the basis of $\mathcal{A}_C$. This produces $h_{ij}$ operators that have both positive and negative eigenvalues. The parent Hamiltonian then has $\ket{\psi_\mathrm{MPS}}$ as an eigenstate with zero eigenvalue but not necessarily as an extremum. We then add small random perturbations supported on $\mathcal{A}_C$ to ensure that the spectrum has GOE-like level statistics.

The parent Hamiltonian embedding provides a benchmark that is qualitatively different from the Shiraishi-Mori approach. By varying $\chi$ and the interaction range $D$ we can tune the entanglement of the scar eigenstate. This allows us to examine the performance of $\sigma$-VQE when preparation of the scar requires some entanglement be generated by the shallow circuit.

To validate that our benchmarks contain a single scar low-entanglement eigenstate amid otherwise thermal eigenstates, we plot the bipartite entanglement entropy of the full spectrum in Fig. \ref{fig:ent}. The bipartite entanglement entropy of each eigenstate $\rho_i$ (with corresponding eigenvalue $E_i$) is defined using a contiguous real-space bipartition into subsystems $A$ and $B$, where for the open-chain Shiraishi--Mori model $A$ and $B$ are the left- and right-hand parts of the chain, while for the parent Hamiltonian ring we cut the ring at a fixed bond and take $A$ and $B$ to be two contiguous halves
\begin{equation}
    \mathcal{S}=-\mathrm{Tr}[\rho_{i}^A\log \rho_{i}^A]
\end{equation}
where the reduced density matrices are found by tracing out the support of partition B, i.e $\rho_{i}^A=\mathrm{Tr}_B[\rho_{i}^{AB}]$.

For the Shiraishi-Mori embedding, the embedded scar is a product state and appears as a clearly isolated eigenstate at $E=0$. For parent Hamiltonian embedding with $N=8$, $D=4$, tuning the bond dimension $\chi$ allows us to increase the scar entanglement. As we do this the scar becomes less separated from the thermal band, providing a control parameter for testing how the performance of $\sigma$-VQE degrades as the entanglement gap is reduced (and thus the shallow ansatz becomes insufficient to prepare mid-spectrum eigenstates).

\FloatBarrier
\section{Results}\label{sec:results}

We benchmark the $\sigma$-VQE using both models outlined in Sec. \ref{sec:scars} with the goal of assessing whether a shallow, hardware-efficient ansatz combined with an energy-targeting, variance-penalized objective can reliably converge to a QMBS when such an atypical low-entanglement eigenstate exists near a chosen target energy. We quantify performance using the cost function, Eq. \ref{eq:cost}, taking $a=b=1/2$. We additionally track the fidelity of the state produced by the circuit, $\ket{\psi_{\theta}}$, with the target scar state, $\ket{\psi_\mathrm{scar}}$
\begin{equation}\label{eq:fidelity}
\mathcal{F}_{\theta}=| \langle {\psi_\mathrm{scar}}| \psi_{\theta} \rangle|^2.
\end{equation}
\subsection{Statevector simulations}
We first benchmark $\sigma$-VQE using exact statevector simulations for numerical evaluation of the cost objective (and its gradients). We then make these simulations more realistic by using the techniques described in \ref{subsec:noise} to simulate the effect of shot noise on trainability. In both cases we assume noiseless circuit execution and measurement. Optimization is performed with ADAM using PSR, and the hardware-efficient ansatz is initialized from $\ket{0}^{\otimes N}$ with small random parameters (of order $10^{-3}$).

\begin{figure}[t]
    \centering
    \includegraphics[width=0.9\columnwidth]{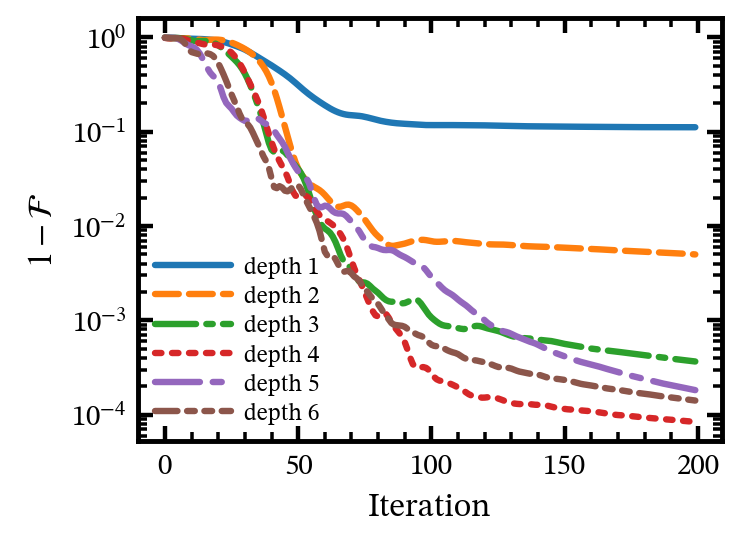}
    \caption{$\sigma$-VQE performance when searching for the SM-embedded non-integrable $N=9$ open XXZ chain. The results shown are from setting the target energy of the cost function to 0, the energy of the scar state. We use ADAM with exact (statevector) evaluation of all expectation values and parameter-shift gradients. We simulate using 1 to 6 ansatz layers. We plot the infidelity ($1-\mathcal{F}$) of the circuit state with the scar state vs iteration of the algorithm. We observe that even a relatively shallow depth of $2$ is enough to achieve significant overlap with the scar state in this setting.}
    \label{fig:depth}
\end{figure}

\begin{figure}[hbt!]
    \centering
    \includegraphics[width=0.4\textwidth]{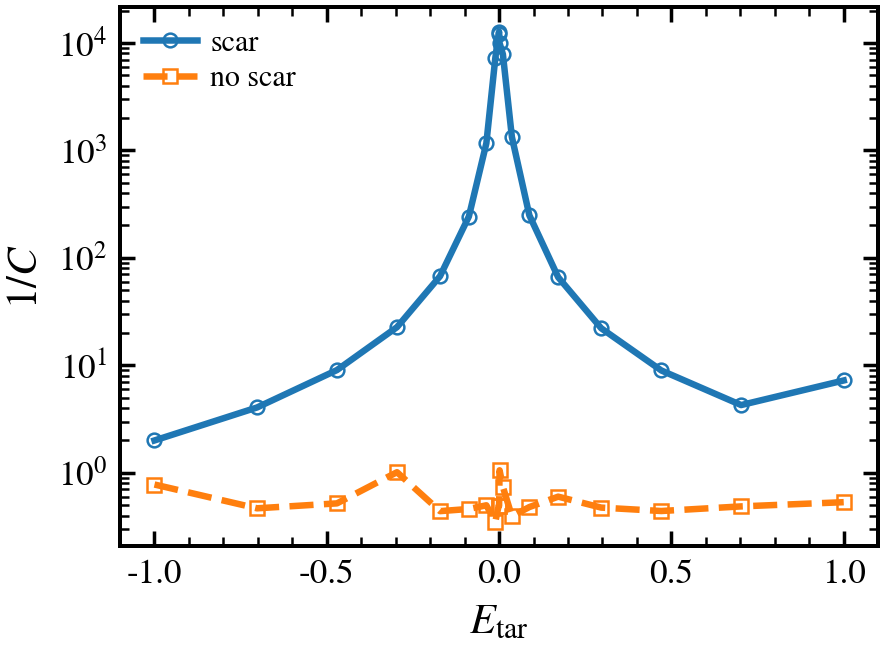}
    \caption{We run the $\sigma$-VQE for the $N=9$ Shiraishi-Mori embedded chaotic XXZ model, now varying the target energy that is set in the cost function, Eq. \ref{eq:cost}. We run the $\sigma$-VQE routine for $300$ iterations with an ansatz depth of 3. We observe that the algorithm is only successful in reducing the cost function significantly when we specifically target the scar. Eigenstates at other target energies require more entanglement to be generated, and therefore are out of reach of the low-depth circuit. We additionally run the VQE for a control case of the chaotic XXZ model without an embedded scar. The shallow ansatz cannot significantly lower the cost function at all in the control case.}
    \label{fig:target}
\end{figure}

Fig. \ref{fig:depth} shows $\sigma$-VQE applied to the $N=9$ Shiraishi–Mori-embedded open XXZ chain taking $E_{\mathrm{tar}}=0$, the eigenvalue of the scar state. We run the algorithm with circuit depths ranging from $1$ to $6$. All other hyperparameters are kept fixed. The main observation is that substantial overlap with the scar is already achievable with low-depth circuits, indicating that the low-entanglement structure of the target state is accessible to the shallow ansatz. The results shown do not suggest a simple monotonic improvement with depth, and we therefore limit ourselves here to the conclusion that low-depth circuits already perform well in this benchmark.
To test whether $\sigma$-VQE succeeds in lowering the cost function only near anomalously low-entanglement eigenstates, we use the same $N=9$ Shiraishi-Mori XXZ Hamiltonian and sweep the target energy, $E_{\mathrm{tar}}$, over a range of values. For each $E_{\mathrm{tar}}$ value we run a $300$ iteration $\sigma$-VQE routine, and we additionally repeat the procedure for a control case of a disordered XXZ chain without an embedded scar. As shown in Fig. \ref{fig:target}, the cost objective drops substantially only when the target is tuned to the scar energy. Away from the scar the optimization largely stalls, consistent with the fact that the shallow ansatz is unable to represent generic mid-spectrum (volume-law entangled) eigenstates. In the no-scar control case, the same ansatz fails to reduce the cost objective significantly across the scanned energies. This energy selectivity is consistent with the design principle of the $\sigma$-VQE. Together, Fig. \ref{fig:depth} and Fig. \ref{fig:target} illustrate the central mechanism exploited for the algorithm. When entanglement generation is intentionally restricted, successful eigenstate targeting becomes selective for those states that are anomalously low-entanglement in an otherwise ETH-obeying system.

These noiseless baseline tests establish functionality in the limit of infinite measurement statistics. We next introduce finite-shot sampling of the cost, keeping the underlying circuit evolution noiseless. In this regime, the cost is evaluated via the unbiased estimators defined in Sec.~\ref{subsec:unbiased}. We reiterate that we reuse each measured bitstring across multiple Pauli strings, with simultaneous reconstruction of multiple expectation values within a qubit-wise commuting group to reduce the number of circuits required for accurate measurement of expectation values. Under independent sampling, the unbiased estimator variance is expected to decrease as $1/S$ due to the central limit theorem, where $S$ is the shot budget per gradient evaluation. Different grouping schemes for Pauli strings primarily affect the prefactor for this decay. A number of schemes for grouping Pauli strings and creating measurement settings $\mathcal{B}$ have been proposed in the literature \cite{hillmich2021decision,huang2021efficient,elben2023randomized,izmaylov2019unitary,verteletskyi2020measurement} in an effort to reduce this estimation variance and the number of circuits required for the execution of NISQ algorithms and beyond. For our purposes we use Qiskit’s default QWC grouping scheme for generating the measurement bases.
\begin{figure*}[hbt!]
\centering
\includegraphics[width=0.45\textwidth]{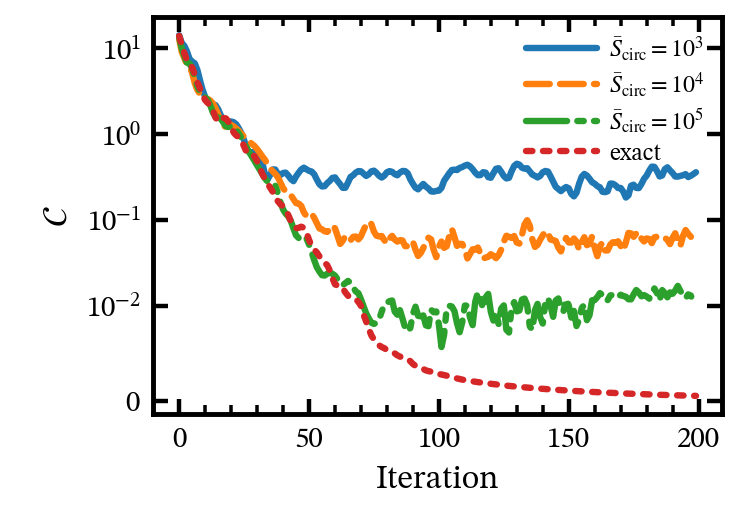}\includegraphics[width=0.45\textwidth]{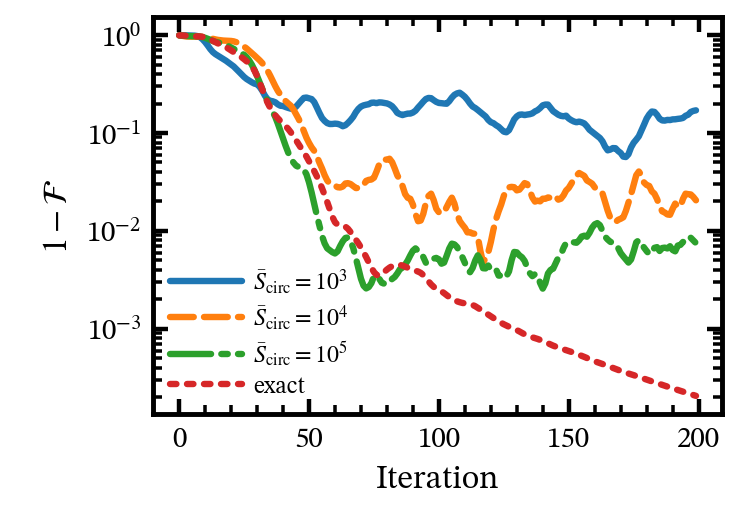}

\caption{
We simulate the \(\sigma\)-VQE with a depth of 3 for the
\(N=9\) open XXZ chain with a Shiraishi--Mori
random-product QMBS state, using average shot budgets
\(\bar S_{\rm circ}\in\{10^3,10^4,10^5\}\) per physical
QWC circuit evaluation. The plotted cost \(\mathcal C\)
is the exact statevector cost evaluated along the finite-shot
optimization trajectories. The parameter updates themselves
are computed using shot-noisy estimates. This probes whether
shot-noisy parameter updates still produce progress in the
underlying noiseless cost landscape.
}
\label{fig:noisy}
\end{figure*}

Fig.~\ref{fig:noisy} shows the same $\sigma$-VQE being run for the
\(N=9\) SM embedded chaotic XXZ model using the QWC group
estimation scheme. We report the finite-shot data in terms of the
average number of shots per physical QWC circuit evaluation,
\(\bar S_{\rm circ}\). This corresponds to a total shot budget
\(S_{\rm eval}=G\bar S_{\rm circ}\), where \(G\) is the number of
QWC measurement groups. In this instance \(G=193\), and the budget
is divided across groups using the 
importance-weighted allocation described in Sec. \ref{subsec:noise}.

For the depth-3 ansatz, \(P=126\), so each parameter-shift rule ADAM
update requires \(2P+1\) estimator
evaluations, giving a total shot count per iteration of 
\((2P+1)G\bar S_{\rm circ}\) not including any additional cost-logging evaluation. We use
\(\bar S_{\rm circ}\in\{10^3,10^4,10^5\}\), where \(10^3\) and
\(10^4\) are typical reference shot counts per
circuit and \(10^5\) is a high-statistics reference.

In the cost panel we plot the exact statevector cost \(\mathcal{C}\) evaluated at the parameters produced by the finite-shot optimization. The parameter updates are computed using shot-noisy estimates, while the displayed cost is evaluated without shot noise, separating progress in the underlying noiseless cost landscape from fluctuations of the finite-shot cost estimator. We stress that the noiseless curve is a reference optimization trajectory, not a lower bound on the infidelity or cost. Finite-shot gradient estimates generate different stochastic optimization paths, and fidelity is not the optimized quantity, so a finite-shot trajectory can pass through points with lower instantaneous infidelity than the noiseless reference at the same iteration. At smaller \(\bar{S}_{\rm circ}\), the larger estimator variance limits the precision with which the cost landscape and gradients are resolved, consistent with the higher apparent optimization floor in the examples shown. We therefore conclude that the unbiased estimator remains able to support \(\sigma\)-VQE optimization under finite-shot sampling, although the total PSR-ADAM circuit overhead remains substantial.

The Shiraishi-Mori benchmark embeds a strict product eigenstate and is therefore ``easy” from an entanglement standpoint. As the qubit register is initialized in a product state and we select small initial circuit parameters, optimization is relatively straightforward. To assess performance beyond product scars, we benchmark $\sigma$-VQE on the parent Hamiltonian (PH) construction, which embeds a matrix product state (MPS) as an exact eigenstate with bond dimension $\chi$ (see Sec. \ref{sec:ph} for details). The bond dimension is a free parameter which allows us to increase the entanglement of the embedded scar, providing a controlled interpolation between product-like and more entangled (yet still sub-thermal) target states. This is visible directly in the exact-spectrum entanglement diagnostics in Fig. \ref{fig:ent}. As $\chi$ increases, the anomalous eigenstate becomes less isolated from the thermal band. Consequently, $\sigma$-VQE is forced into a regime where the circuit ansatz must generate enough structure to represent the scar, while still remaining too entanglement-limited to approximate generic mid-spectrum eigenstates at nearby energies.

\begin{figure*}[hbt!]
\centering
\includegraphics[width=0.45\textwidth]{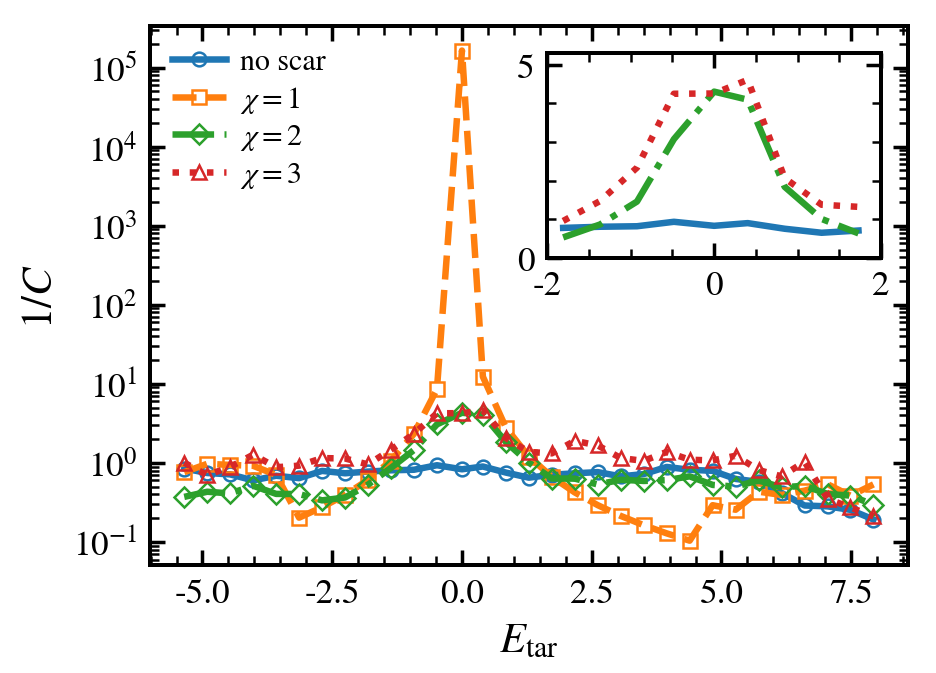}\includegraphics[width=0.45\textwidth]{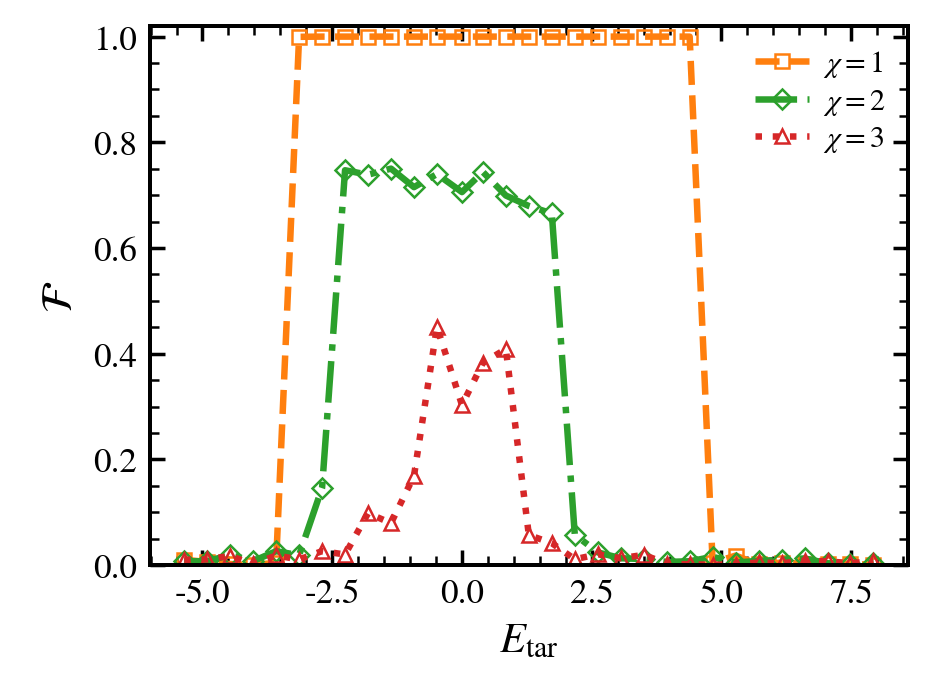}
\caption{$\sigma$-VQE performance when searching for a random matrix product state QMBS that is embedded via the parent Hamiltonian method taking $N=8$, $\chi=1,2,3$ and $D=4$ together with a no-scar control Hamiltonian. For each target energy $E_{\mathrm{tar}}$, we run a single depth-$4$ hardware-efficient ansatz optimization using 300 iterations of PSR gradient evaluations. We plot inverse cost and fidelity to the embedded MPS scar versus $E_{\mathrm{tar}}$. Broad fidelity plateaus compared to a narrower inverse-cost peak indicate that the variational search can still find substantial scar overlap even while the energy detuning is penalized by the cost function.}
\label{fig:MPS}
\end{figure*}

We benchmark $\sigma$-VQE using the parent Hamiltonian construction in Fig. \ref{fig:MPS}. We use $N=8$ qubits, interaction range $D=4$, and a depth-$4$ hardware-efficient ansatz. For each $\chi\in\{1,2,3\}$ and at each target energy $E_{\mathrm{tar}}$, we run 300 iterations of the $\sigma$-VQE routine using PSR gradient evaluations. The same procedure is repeated for a control Hamiltonian with similar structure in which no scar is embedded. We plot both the inverse cost and the fidelity to the known scar state (when one is present). The inverse cost peaks around $E_{\mathrm{tar}}=0$, the energy of the embedded scar state, reflecting the explicit energy-targeting structure of the $\sigma$-VQE objective function. The fidelity exhibits broader plateaus instead of peaks as a function of $E_{\mathrm{tar}}$ than the inverse of the cost function. The optimization can converge to states with appreciable overlap with the embedded MPS scar while $E_{\mathrm{tar}}$ is detuned from the exact eigenvalue of the scar despite the cost function itself being penalized for that detuning. This is consistent with the hypothesis of scar states being particularly attractive low-entanglement solutions for the variational search with shallow circuits. The inverse cost peak is therefore a more suitable diagnostic of the correct energy window. As $\chi$ increases, $\sigma$-VQE performance as a high-fidelity state preparation protocol degrades. Increasing the MPS bond dimension raises the entanglement required to represent the scar and reduces its separation from the thermal band, narrowing the regime where a shallow ansatz can represent the scar while still failing to select for typical eigenstates. The no-scar control provides the corresponding null test. In the absence of an anomalous low-entanglement eigenstate, the same shallow variational search does not produce a comparable signature with the cost function.

The parent Hamiltonian results support two practical conclusions relevant to scar-hunting in less engineered models. First, the $\sigma$-VQE need not be restricted to product-like scars. It can identify and prepare moderately entangled scars so long as they remain atypical relative to the surrounding thermal spectrum. Additionally, the observed fidelity plateaus suggest an experimentally useful workflow when the scar energy is not known a priori. A coarse sweep in $E_{\mathrm{tar}}$ can be used to locate candidate windows via an inverse-cost peak, while the broad overlap region indicates that the algorithm can still converge to the relevant low-entanglement structure even before the target energy is perfectly tuned. 

We note some further improvements can be made to our strategy if we wish to maximize fidelity with the scar state. While we restrict Fig. \ref{fig:MPS} to single-run optimization to characterize baseline behavior, standard variational-algorithm strategies may improve robustness in higher $\chi$ regimes. Examples include restart strategies with different initial circuit parameters to mitigate optimizer sensitivity in rugged landscapes combined with the use of cheaper initial optimizers such as SPSA when one wishes to reduce per-iteration measurement overhead to explore many initial positions in the optimization landscape quickly. More broadly, the VQE literature contains a range of practical techniques that are known to affect convergence. Comparative studies show strong dependence on the classical optimizer, with gradient-based methods often outperforming common derivative-free choices when gradients are accessible \cite{ComparativeOptimizersVQE2025,Lavrijsen2020ClassicalOptimizers}. Geometry-aware updates such as the quantum natural gradient can further improve performance \cite{Stokes2020QNG}. Likewise, circuit/parameter initialization strongly affects trainability. A given random initialization can place the optimization on barren plateaus \cite{McClean2018BarrenPlateaus}, motivating strategies such as identity-block initialization \cite{Grant2019IdentityInit}. Additionally, optimizing the circuit structure itself can improve outcomes \cite{Ostaszewski2021StructureOpt}.

Finally, as the estimator variance depends on how shots are distributed across measurement bases, variance-aware shot allocation can accelerate convergence at fixed measurement budget. Shot-adaptive approaches allocate shots based on gradient/estimator variance to aid estimation \cite{Kubler2020iCANS,gu2021adaptive,OperatorSampling2020,Zhu2023ShotAssign}. In addition, reinforcement-learning methods for shot assignment have also been proposed \cite{AIDrivenShotReduction2024}. In the present context, such methods are best viewed as ways to improve performance at fixed shot counts, rather than qualitatively changing the mechanism behind $\sigma$-VQE.

\FloatBarrier
\subsection{Hardware experiment}

We demonstrate a proof-of-principle execution of $\sigma$-VQE on superconducting hardware. Using several IBM Quantum Platform trial allocations, we obtained brief access (50 minutes total) to the IBM Fez backend. 
\begin{figure*}[hbt!]
\centering
\includegraphics[width=0.45\textwidth]{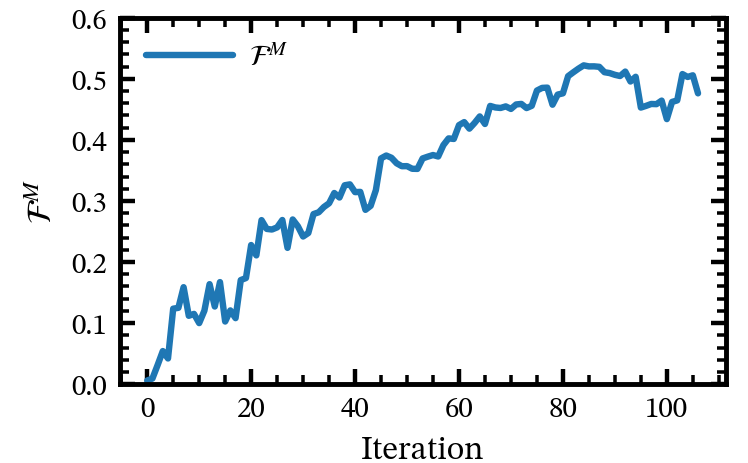}\includegraphics[width=0.45\textwidth]{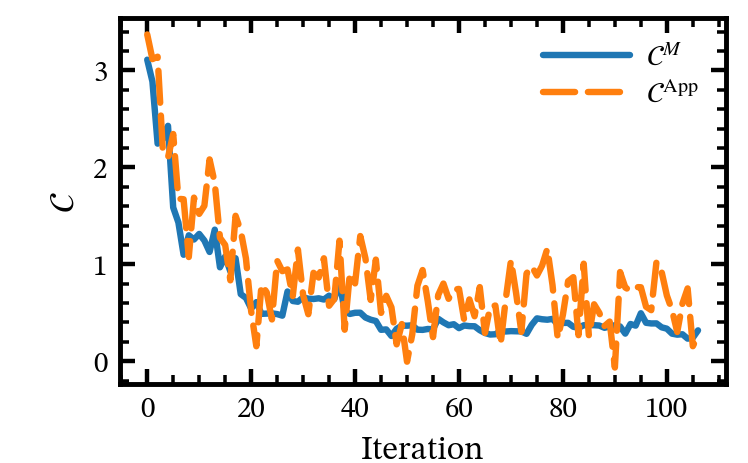}

\caption{
Hardware implementation of $\sigma$-VQE for the $N=3$ SM-embedded XXZ chain on IBM Fez (Heron r2), using a depth-2 hardware-efficient ansatz and SPSA optimization. We plot the on-device proxy cost $C^{\mathrm{App}}$ computed from shifted evaluations used by SPSA together with an offline reconstructed ``model” cost $C^M$ and model fidelity $\mathcal{F}^M$ obtained by replaying the hardware parameter trajectory in a noiseless statevector simulation. We used dynamical decoupling on idle qubits and Pauli twirling on two-qubit gates and readout as error mitigation. Total QPU usage was 50 minutes.
}
\label{fig:experiment}
\end{figure*}
We restrict to error mitigation techniques that do not multiply sampling cost by large constant factors: we enable dynamical decoupling sequences on idle qubits and apply Pauli twirling to two-qubit gates and measurement. In contrast, we do not use methods such as probabilistic error cancellation, error amplification, or zero-noise extrapolation \cite{cai2023quantum,Kandala2019ErrorMitigation}. While powerful, they each significantly increase circuit volume and/or require repeated evaluations at multiple effective noise levels, which is incompatible with the present shot-limited demonstration and with shallow-budget access regimes.

In implementing our estimator described in Subsec. \ref{subsec:unbiased}, we utilized the Qiskit Sampler primitive rather than the Estimator primitive. Sampler returns raw shot bitstrings for each circuit instance, which was required by our reuse of each bitstring across multiple Pauli strings within the same QWC measurement group. Estimator aggregates observables internally and therefore discards the bitstring-level correlations exploited by our post-processing. Use of the Sampler primitive also limits the available error mitigation techniques in Qiskit.

The limited shot budget prevented us from logging the cost function at each iteration of the optimizer. Since SPSA only requires the shifted evaluations $C(\theta_t^{(+)})$ and $C(\theta_t^{(-)})$ to construct its gradient estimate, we devoted the available shots entirely to these evaluations and monitored progress using an on-device proxy that required no additional circuit executions
\begin{equation}\label{eq:capp}
    C_k^{\mathrm{App}}=\frac{C(\theta_k^{(+)})+C(\theta_k^{(-)})}{2},
\end{equation}
where $C(\theta_k^{(\pm)})$ are the shifted cost function evaluations used for the SPSA optimizer as defined in \eqref{eq:shifted}. 

To maximize robustness, we intentionally target a minimal model, the $N=3$ Shiraishi–Mori embedded chaotic XXZ model with a depth-$2$ hardware-efficient ansatz. The low depth reduced the two-qubit gate count, idle time, and exposure to readout noise. We additionally transpiled the logical chain onto a connected set of physical qubits with the lowest available two-qubit error rates. Deeper circuits and larger $N$ degrade the optimization signal at fixed shots. Emulation of the noisy hardware suggested that instances beyond $N=3$ were not reliably trainable without stronger mitigation (App. \ref{App:hardware}).

 The hardware results are shown in Fig. \ref{fig:experiment}. We used 5,000 shots per SPSA iteration ($535{,}000$ total). The on-device proxy $C_k^{\mathrm{App}}$ shows a clear downward trend, indicating that—even under restricted sampling and lightweight mitigation—the $\sigma$-VQE objective retains a usable optimization signal on hardware. To interpret this trajectory beyond a noisy proxy, we also perform an offline ``model replay." We take the parameter sequence $\{\theta_t\}$ produced by the on-device SPSA loop and evaluate the corresponding ``model" cost $C_k^{M}$ and ``model" fidelity $\mathcal{F}_M$ in a noiseless statevector simulation (where the embedded scar is known numerically). The replay confirms that the parameter trajectory learned on the noisy hardware corresponds to improved model cost and increased overlap with the scar. The significance of the hardware run is not the absolute system size, but that the combination of a shot-frugal optimizer, bitstring-level post-processing for nonlinear objectives, and lightweight mitigation suffices to obtain an on-device optimization signal.

\FloatBarrier

\section{Concluding remarks}
In this work we benchmark a hybrid quantum–classical protocol, $\sigma$-VQE, for uncovering quantum many-body scar eigenstates. The method combines a nonlinear, energy-targeting objective with a variance penalty, alongside a shallow hardware-efficient circuit ansatz that restricts accessible entanglement to bias optimization toward low-entanglement eigenstates \cite{Cenedese2025}. The resource overhead of the algorithm is reduced by a classical shadow-motivated qubit-wise commuting (QWC) measurement strategy together with an unbiased finite-shot estimator for the nonlinear cost function, ensuring that finite-sampling effects do not systematically bias the variance-related terms. 

Across noiseless simulations, $\sigma$-VQE rapidly converges for Shiraishi–Mori projector-embedded product scars, while parent Hamiltonian MPS embeddings provide a controlled view of failure modes as the embedded scar entanglement increases towards typical thermal eigenstates in the same energy window. Finally, a proof-of-principle superconducting-hardware execution (Fig. \ref{fig:experiment}) demonstrates end-to-end feasibility under tight shot constraints using a shot-frugal optimizer, bitstring-level post-processing, and lightweight mitigation.

These results support interpreting $\sigma$-VQE as a diagnostic for scarred structure near a chosen target energy density. If a low-entanglement scar exists near the target energy, the shallow ansatz can represent it and the nonlinear objective supports convergence, whereas the same circuit family fails to realize generic mid-spectrum eigenstates. The results additionally show that $\sigma$-VQE acts as a state-preparation primitive for initializing states with significant overlap with nonthermal states for subsequent dynamical studies, such as probing anomalous relaxation \cite{bernien2017probing,turner2018quantum}. Many applications do not require high-fidelity scar eigenstate preparation. Preparing a state with appreciable overlap onto a scar, as seen in the MPS case, is sufficient to exhibit long-lived coherent dynamics, consistent with the wider scar literature \cite{bernien2017probing,turner2018quantum,hummel2023genuine,bluvstein2021controlling}. This perspective also aligns with proposed scar-enabled applications such as robust sensing and metrological enhancement \cite{dooley2021robust,dooley2023entanglement}.

$\sigma$-VQE fits within the broader class of variational excited-state methods \cite{higgott2019variational,nakanishi2019subspace,mcclean2017hybrid,gocho2023excited}. As with all VQEs, its practical scaling is constrained by the growth of measurement overhead and the fragility of the optimization signal under finite-shot, noisy execution. In our setting, lightweight mitigation is sufficient for a proof-of-principle hardware demonstration, but scaling to larger systems on current devices will likely require either substantially higher shot budgets or stronger error mitigation to recover a useful optimization signal \cite{Kandala2019ErrorMitigation,Kurita2023SynergeticQEM,cai2023quantum,Wang2024EMTrainability}. Improving measurement efficiency to counteract the growth of Pauli terms (such as through grouping, derandomization, parallelization, and variance-aware shot allocation \cite{verteletskyi2020measurement,huang2020predicting,huang2021efficient,gu2021adaptive}) and better device-level noise performance will therefore be central to extending $\sigma$-VQE beyond the small-system regime.

\begin{acknowledgments}
The authors thank Shane Dooley and Phillip C. Burke for insightful discussions on the models explored in this work. Eoin Carolan would particularly like to thank Chloe Smith and Jack Rivington for conversations on the broader scope of the project near its outset.
EC, GC, and GB acknowledge financial support from the ``National Centre for HPC, Big Data and Quantum Computing'', Spoke 10, Project CN00000013
QUOVADIS and from INFN through the project QUANTUM.
NK and GC acknowledge the Spanish State Research Agency through the María de Maeztu project CEX2021-001164-M and the COQUSY project PID2022-140506NB-C21 and -C22, all funded by MCIU/AEI/10.13039/501100011033.
NK further acknowledges the QuantCom project CNS2024-154720. GC further acknowledges the INFOLANET project PID2022-139409NB-I00 and the QuantERA QNet project PCI2024-153410, funded by MICIU/AEI/10.13039/501100011033 and by ERDF, EU.

We acknowledge the use of IBM Quantum services for this work. The views expressed are those of the authors, and do not reflect the official policy or position of IBM or the IBM Quantum team.
The data presented in this work are available upon request.
\end{acknowledgments}
\appendix

\section{Hardware emulation}\label{App:hardware}

\begin{figure*}[hbt!]
\centering
~~\includegraphics[width=0.48\textwidth]{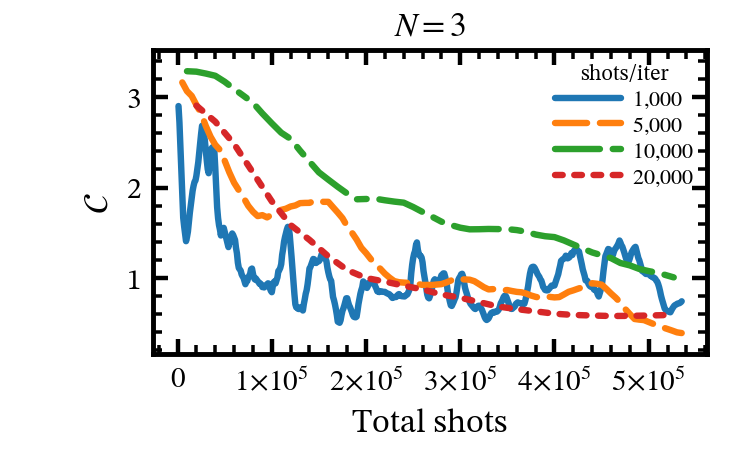}~~\includegraphics[width=0.48\textwidth]{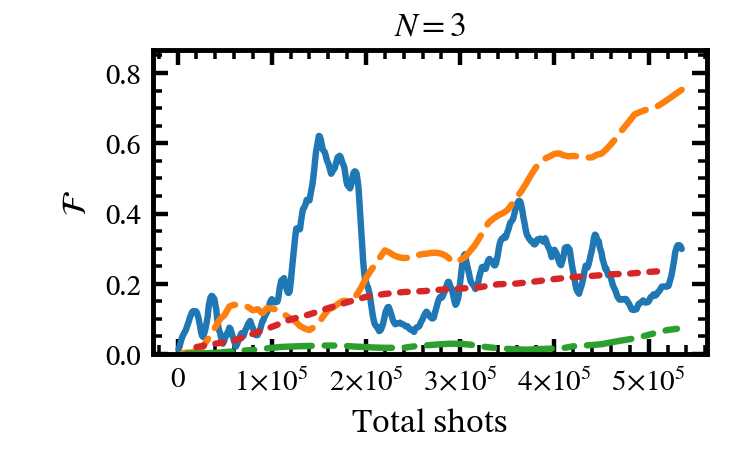}\\~~\includegraphics[width=0.48\textwidth]{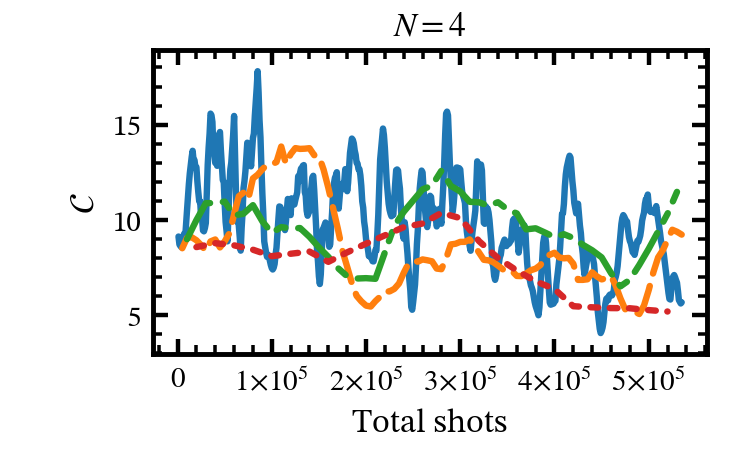}~~\includegraphics[width=0.48\textwidth]{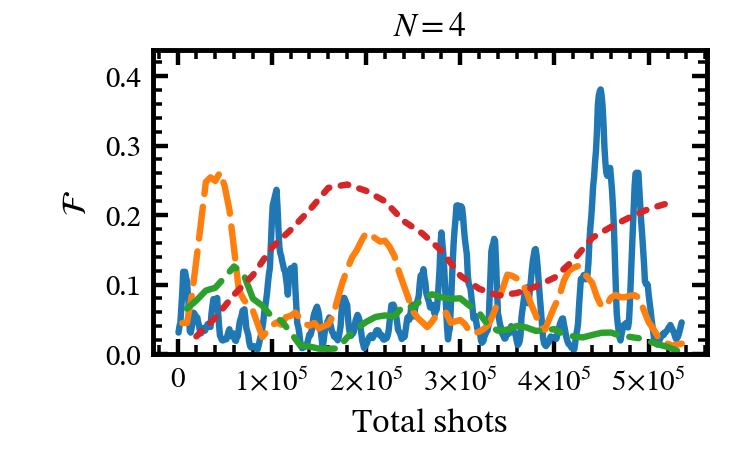}

\caption{The results of our noisy emulation using $535{,}000$ total shots (the amount accessed during the hardware run). We use SPSA and a depth-2 hardware-efficient ansatz. Finite-shot outcomes were generated via multinomial sampling from the noisy output distributions. We compare system sizes of $N=3$ and $N=4$ while using different budgets for shots per SPSA iteration $S$. For $N=3$, optimization remains viable while for $N=4$ the noise model prevents reliable convergence within the same shot budget.}
\label{fig:emulatedexperiment}
\end{figure*}

We used Qiskit Aer to emulate $\sigma$-VQE runs under a device-calibrated noise model for IBM Fez. For a chosen physical-qubit chain, we transpiled the variational circuits onto the Fez device layout and constructed a noise model from the backend calibration data, including gate error rates and durations, per-qubit T1/T2 times, and readout assignment errors. This emulator captures dominant stochastic error channels at the level provided by the calibration snapshot, but it does not include several effects that can be relevant on hardware, such as coherent over/under-rotations and miscalibrations, drift, and crosstalk. The emulation results should therefore be interpreted as an optimistic proxy for trainability rather than a faithful predictor of hardware performance.

We fixed a total budget of $535{,}000$ shots and used SPSA together with a depth-2 hardware-efficient ansatz, reflecting the shot- and depth-frugal regime targeted in the hardware demonstration. We searched over a grid of shots per iteration of the SPSA optimizer ($S=10^3,5\times10^3,10^4,2\times 10^4$) which trades estimator variance against the number of optimizer iterations available within the fixed total budget. For each choice of $S$ we generated the requisite number of bitstring outcomes via the multinomial sampling procedure described in Subsec. \ref{subsec:noise}. These bitstrings were then used to estimate the shifted cost evaluations for the SPSA gradient approximation. We plot the cost and fidelity with the scar state using access to the full simulated density matrix of the circuit.

The results are shown in Fig. \ref{fig:emulatedexperiment}. For $N=3$ optimization remained viable within the fixed budget. We find that allocating fewer shots per iteration (hence allowing more iterations) generally improved the final outcome until the SPSA updates became dominated by shot noise. In contrast, for $N=4$ the same budget and circuit depth did not yield reliable convergence under the Fez-calibrated noise model. This supports the design choice used for the hardware run. Within the available shot budget and without stronger mitigation, increasing N rapidly degrades the optimization signal, and larger instances are unlikely to be trainable in the same regime.

\FloatBarrier

\bibliography{bibliography}

\end{document}